\documentclass[aps,prb,twocolumn,groupedaddress]{revtex4}

\usepackage{amsmath}
\usepackage{epsfig}
\usepackage{graphics}

\newcommand{\daka}[1]{{#1}^{\dagger}}
\def\ba{\begin{array}}
\def\ea{\end{array}}

\newcommand{\tmpnote}[1]%
   {\begingroup{\it (FIXME: #1)}\endgroup}

\begin{document}


\title{State-dependent impedance of a
strongly coupled oscillator-qubit system}


\author{Teemu Ojanen}
\email[Correspondence to ]{teemuo@boojum.hut.fi}

\affiliation{Low Temperature Laboratory, P.O. Box 2200, FIN-02015
HUT, Finland}

\author{Tero T. Heikkil\"a}

\affiliation{Low Temperature Laboratory, P.O. Box 2200, FIN-02015
HUT, Finland}

\affiliation{Department of Physics and Astronomy, University of
Basel, Klingelbergstr. 82, CH-4056 Basel, Switzerland}


\date{\today}

\begin{abstract}
We investigate the measurements of two-state quantum systems
(qubits) at finite temperatures using a resonant harmonic
oscillator as a quantum probe. The reduced density matrix and
oscillator correlators are calculated by a scheme combining
numerical methods with an analytical perturbation theory.
Correlators provide us information about the system impedance,
which depends on the qubit state. We show in detail how this
property can be exploited in the qubit measurement.

\end{abstract}


\maketitle


\section{Introduction}

The growing interest in quantum information theory and the rapid
development in nanotechnology have resulted in an extensive study of
quantum two-state systems (qubits) and the quantum measurement
theory. The measurement theory of frequency independent detectors
has been studied in detail in recent
years.\cite{bragisky,averin,korotkov,clerk} Motivated by the new
directions of research, we investigate possibilities to use a
harmonic oscillator as a generic frequency dependent measuring
device of a qubit.

\begin{figure}[h]
\centering
\begin{picture}(200,80)
\put(50,0){\includegraphics[width=0.4\columnwidth]{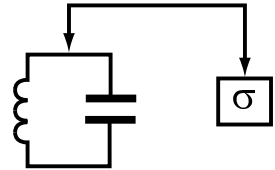}}
\end{picture}
\caption{Possible realization of the studied system. The resonator
circuit is coupled to the qubit represented by $\sigma$. In
practice $\sigma$ could be for example a Cooper-pair box with a
capacitive coupling to the oscillator.} \label{skeema}
\end{figure}

Our system under study consists of a qubit coupled resonantly to a
harmonic oscillator (Fig.~\ref{skeema}), both coupled to a bosonic
heat bath at a finite temperature. This model has a wide range of
applications in solid-state physics as well as in quantum optics
and has raised considerable attention
lately.\cite{blais,schuster,girvin,ilichev,smirnov,makhlin,chiorescu,
wallraff} In the literature there exists various propositions to
realize this system. In solid-state physics the harmonic
oscillator is realized by a resonator circuit and the qubit, for
example, by a Josephson charge or flux qubit. The heat bath
corresponds to the electromagnetic environment of the circuit. The
connection of these systems to cavity QED has been explained and
studied in Refs.~\onlinecite{rau, blais}.

Recently, also experimental studies of solid-state realizations of
 cavity QED have become accessible. Wallraff \emph{et al.} \cite{wallraff}
performed an experiment where they successfully managed to couple a
transmission line resonator with a Cooper pair box acting as a
qubit. They found a clear evidence of a quantum entanglement between
the resonator and the qubit by measuring the vacuum Rabi mode
splitting. Chiorescu  and coworkers\cite{chiorescu} realized the
oscillator+qubit system by coupling a Josephson flux qubit to a
SQUID. In their experiment the SQUID behaved like a harmonic
oscillator and was used as a measuring device of the qubit. The
entangled states of the oscillator-qubit system could be generated
and controlled through the use of microwave spectroscopy. The fact
that the qubit is fixed on the same chip as the resonator allows
circuit QED systems to explore the strong coupling regime of cavity
QED, difficult to reach in quantum optics realizations
\cite{makhlin}.

In the light of these recent experiments, we examine how the qubit
information can be extracted from the oscillator measurement.
 Oscillator correlators reveal the entanglement between the qubit and
 the oscillator. The correlator information can be linked to
 experimentally accessible quantities. We propose a scheme to
 extract information about the
state of the qubit by probing the impedance of the oscillator.
This can be realized by a transmission measurement. As a result of
such an experiment, one can resolve the diagonal state of the
qubit. Such transmission measurements have been recently exploited
in studying the solid-state cavity QED \cite{wallraff}.

Our paper is organized as follows. In Sec. II we detail the
general formalism for calculating the correlators and density
matrices. We also explain the basics of the susceptibility
measurement, and discuss the experimental realization of this
measurement. In Sec. III we study the susceptibility in various
circumstances and how the state of the qubit can be read from the
susceptibility information. Finally, we summarize and discuss the
susceptibility measurement in Sec. IV.

\section{General formalism}

Our starting point  is the Jaynes-Cummings Hamiltonian, which
describes a qubit coupled to a harmonic oscillator in the absence
of dissipation:
\begin{equation}\label{jc}
H_{\mathrm{JC}}=\hbar\omega_0\,\daka{a}
a-\frac{\hbar\omega_{\mathrm{qb}}}{2}\sigma_z+\frac{\hbar
g}{2}\sigma_x(a+a^{\dagger}).
\end{equation}
Operator $\sigma$/$a$ operates on the qubit/oscillator degrees of
freedom. The operators $a$ and $a^{\dagger}$ obey the usual
bosonic commutation relations. We assume the oscillator in
resonance with the qubit, $\omega_0=\omega_{\mathrm{qb}}$. In such
a case the excitation spectrum is doubly degenerate if
oscillator-qubit coupling $g$ vanishes. A finite coupling
$g\ll\omega_0$ slightly lifts the degeneracy of the spetrum
leading to the Rabi splitting. The properties of the spectrum of
(\ref{jc}) are analyzed in detail in Ref.~\onlinecite{rau}.

The total Hamiltonian under study is
\begin{equation}\label{total}
H=H_{\mathrm{JC}}+H_{B}^{\mathrm{qb}}+H_{B}^{\mathrm{osc}}+H_{\mathrm{int}}^{\mathrm{qb}}
+H_{\mathrm{int}}^{\mathrm{osc}},
\end{equation}
where $H_{B}^{\mathrm{qb}}=\Sigma_i\hbar\omega_i\,\daka{b_i}b_i$ and
$H_{B}^{\mathrm{qb}}=\Sigma_j\hbar\omega_j\,\daka{c_i}c_i$ describe
the environments of the qubit and the oscillator, respectively. In
addition,
$H_{\mathrm{int}}^{\mathrm{qb}}=\sigma_x\Sigma_ig_i(b_i+\daka{b_i})$
and
$H_{\mathrm{int}}^{\mathrm{osc}}=(\daka{a}+a)\Sigma_jg_j(c_j+\daka{c_j})$
couple the system to the environment. The chosen bath model is a
popular choice for its conceptual and technical simplicity
\cite{caldeira}.

Our strategy is to calculate the reduced density matrix of the
oscillator-qubit system and the oscillator correlators by
numerically diagonalizing the qubit+oscillator system and taking
the environment into account perturbatively. Only the lowest-order
contributions of the perturbation series are assumed to be
significant. However, to obtain a finite linewidth of energy
levels of the oscillator+qubit system, the perturbation series
must be analytically resummed to the infinite order.

\subsection{Perturbation theory}
We apply the perturbation theory for an open system, developed,
for example, in Refs.~\onlinecite{feynman,rammer,schoeller}. The
time evolution of the density matrix involves forward and backward
propagators which get coupled after tracing out the environmental
degrees of freedom. Given a complete set of energy eigenstates of
a quantum system, the evolution of the density matrix can be
expanded in this basis as
\begin{equation}\label{evol}
 f_{n_1n_2}(t)=\sum_{n'_1,n'_2}\Pi^{n1\,n'_1}_{n_2\,n'_2}(t,t')\,f_{
 n'_1n'_2}(t'),
\end{equation}
where $\Pi^{n1\,n'_1}_{n_2\,n'_2}(t,t')$ is the reduced propagator
and $f_{n_1n_2}(t')$ is the initial reduced density matrix. In the
graphical language the propagator is the sum of all Feynman
diagrams with the given states at each end. The Feynman rules for
an open system, particularly for a quadratic environment with a
bilinear coupling term are described in
Refs.~\onlinecite{feynman,rammer}.

\begin{figure}[h]
\centering
\begin{picture}(200,170)
\put(0,160){(a)}
\put(15,90){\includegraphics[width=0.8\columnwidth]{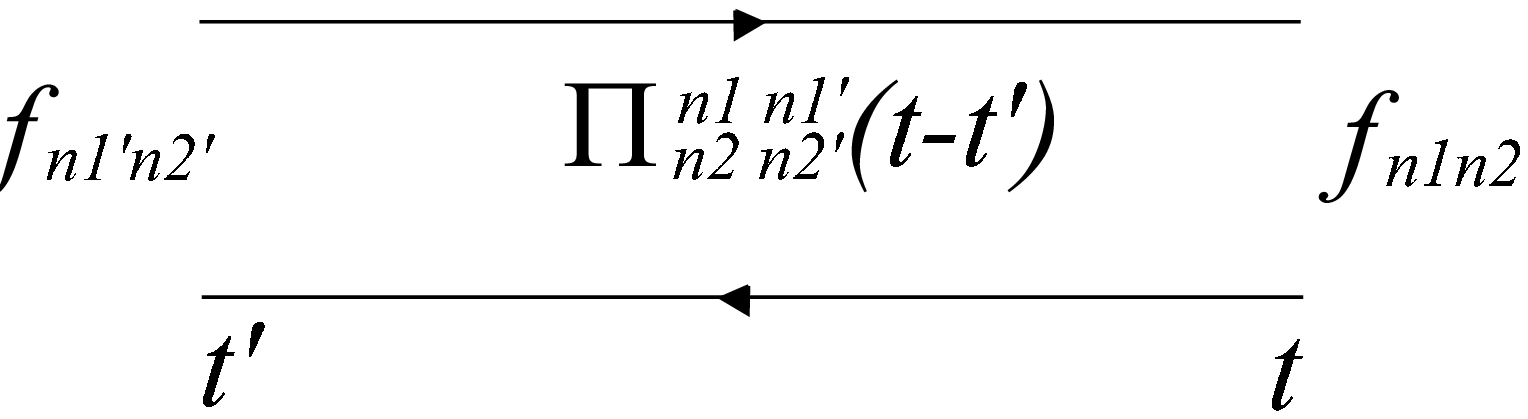}}
\put(0,75){(b)}
\put(0,-20){\includegraphics[width=0.91\columnwidth]{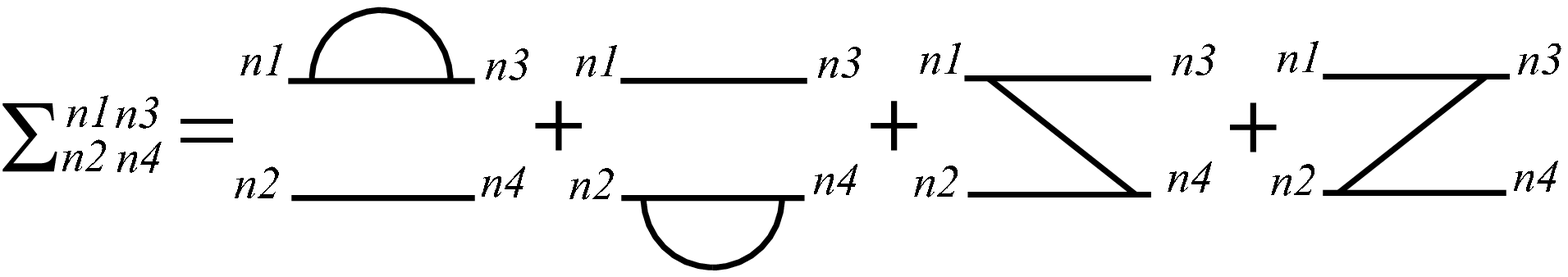}}
\end{picture}
\caption{(a) Propagator evolves the initial state specified by the
density matrix $f_{n1'n2'}$. (b) Second-order contributions to the
irreducible self-energy $\Sigma$.} \label{propaa}
\end{figure}

Supposing that at the initial time the density matrix of the
system+environment is of a tensor product form, the perturbation
theory is significantly simplified. In the frequency space the
propagator is defined as
\begin{equation}\label{muun}
\Pi^{n1\,n'_1}_{n_2\,n'_2}(\omega)=\int_0^{\infty}e^{i\omega(t-t')}\Pi^{n1\,n'_1}_{n_2\,n'_2}(t,t')\,d(t-t').
\end{equation}
The full propagator is described by the Dyson equation
\begin{align}\label{dyson}
\Pi^{n1\,n'_1}_{n_2\,n'_2}(\omega)=\Pi^0\,^{n1\,n'_1}_{n_2\,n'_2}&(\omega)
+\nonumber\\
&\Pi^{n1\,s}_{n_2\,r}(\omega)\,\Sigma^{s\,p}_{r\,l}(\omega)\,\Pi^0\,^{p\,n'_1}_{l\,n'_2}(\omega),
\end{align}
where $\Sigma^{s\,p}_{r\,l}(\omega)$ is the irreducible
self-energy and $\Pi^0\,^{p\,n'_1}_{l\,n'_2}(\omega)$ is the free
propagator. Summation over the repeated indices is implied. The
free propagator elements are determined by the energy eigenvalues
of the reduced system alone:
\begin{equation}\label{free}
\Pi^0\,^{n1\,n'_1}_{n_2\,n'_2}(\omega)=\frac{i}{(E_{n_2}-E_{n_1})/\hbar+\omega}\delta_{n_1n'_1}
\delta_{n_2n'_2}.
\end{equation}
 In our case, the energies in Eq.~(\ref{free}) are the exact energy eigenvalues of
the oscillator+qubit system. The perturbation theory is needed
only in taking the external baths into account. After calculating
$\Sigma^{s\,p}_{r\,l}(\omega)$ to the desired order, the exact
propagator can be solved from Eq.~(\ref{dyson}). In our work we
only take into account the second-order contributions to
$\Sigma^{s\,p}_{r\,l}(\omega)$, see Fig.~\ref{propaa} (b). For
example, the first of the self-energy diagrams represents the
expression
\begin{align}\label{example}
\sum_n\int_{-\infty}^{\infty}&\frac
{J(\omega')f(\omega')d\omega'}{\omega+(E_{n2}-E_{n})/\hbar+\omega'+i\eta}\times\nonumber\\
&\times \langle n|x|n1\rangle\langle
n3|x|n\rangle\delta_{n2\,n4},\nonumber
\end{align}
where $J(\omega')$ is the spectral density of the bath and
$f(\omega')$ is the Bose distribution function. The spectral
density is chosen to produce an ohmic damping.


Correlators $\langle A(t_2)B(t_1)\rangle$ contain important
information about the system. Generally, correlators are functions
of the two time variables $t_1$ and $t_2$, but in a steady state
they depend only on the relative time $t_2-t_1$ due to the
temporal invariance. In this paper we make use of a mixed
representation where correlators depend on one time variable
$t=\mathrm{min}\,(t_1,t_2)$ and one frequency variable $\omega$:
\begin{equation}\label{mix}
D_{AB}(\omega,t)= \int_0^{\infty}e^{i\omega|t_2-t_1|}\langle
A(t_2)B(t_1)\rangle\,d|t_2-t_1|.
\end{equation}
If the system is in a steady state, $D_{AB}(\omega,t)$ is
independent of $t$. The reason for considering correlators of the
type $D_{AB}(\omega,t)$ is that in Sec.~\ref{results} we study the
temporal evolution of the quasistatic susceptibility of the
oscillator+qubit system.

\begin{figure}[h]
\centering
\begin{picture}(200,100)
\put(-15,20){\includegraphics[width=0.9\columnwidth]{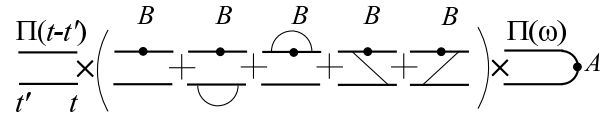}}
\end{picture}
\caption{Diagrammatic representation of the correlator
(\ref{mix}). First the initial state develops to the moment $t$,
then follows external vertex $B$ and the vertex corrections. The
last step is the propagation in the frequency space to the other
external vertex $A$.} \label{mixkorr}
\end{figure}

Correlators cannot be calculated from the reduced propagator and
the initial density matrix alone. The diagrammatic method of
calculating correlators in an open system is presented in
Refs.~\onlinecite{kack,schoeller}. To calculate the correlator
(\ref{mix}), we first need to solve the temporal evolution of the
reduced density matrix from the initial time $t'$ to $t=t_1$
(assuming $t_1<t_2$). This task can be performed with the help of
$\Pi^{n1\,n1'}_{n_2\,n2'}(t,t')$. Next we have to calculate the
vertex corrections in the first external vertex $B$, and propagate
the system with $\Pi^{n1\,n1'}_{n_2\,n2'}(\omega)$ to the other
vertex $A$. Correlator (\ref{mix}) is schematically illustrated in
Fig.~\ref{mixkorr}.

In a stationary state one further simplification occurs. Quantum
regression theorem \cite{cohen-tannoudji} allows one to calculate
the correlator (\ref{mix}) with the propagator and the initial
density matrix alone:
\begin{equation}\label{kor}
D_{AB}(\omega)=
f_{n'_1n'_2}^0\Pi^{n1\,m}_{n_2\,n2'}(\omega)\,A_{n_1n_2}B_{m\,n'_1}.
\end{equation}
In the diagrammatic language this means that the vertex
corrections in the external vertex $B$ cancel out. When
calculating the time-dependent correlator (\ref{mix}) the vertex
corrections must be included, see Fig.~\ref{mixkorr}. Since we
calculated the self-energy to the second order, we take also the
vertex corrections into account to the second order.

\subsection{Numerical scheme}
For computing the propagator, the energy eigenvalues and eigenstates
of the oscillator-qubit system are calculated numerically. This is
done by taking into account only the $N$ lowest energy eigenstates
of the oscillator. The lower the temperature with respect to
$\hbar\omega_0$, the less states $N$ is needed. The propagator
contains $16N^4$ terms. Fortunately, the interaction with the bath
couples only states of the reduced system differing roughly by
$\hbar\omega_0$ in energy\cite{rau}. For this reason, the effective
number of propagator elements is proportional to $N^2$.

The reduced density matrix and correlators can now be computed
from Eqs. (\ref{evol}) and (\ref{kor}) (see also
Fig.~\ref{propaa}).

\subsection{Oscillator measurement}

In studying the effects of the qubit on the oscillator we need to
identify convenient measurable quantities. The idea is to analyze
how, by measuring oscillator observables, we can extract information
about the qubit. In a steady-state situation the cavity
susceptibility of cavity QED is defined as
\begin{equation}\label{suskis}
\chi_x(\omega)=i\int_0^{\infty}e^{i\omega
t}\langle\left[x(t),x(0)\right]\rangle dt,
\end{equation}
where $x=(a+a^{\dagger})$.
The direct analogue of the x-susceptibility (\ref{suskis}) in
solid-state applications is the $\phi$-susceptibility of a resonator
circuit. The resonator circuit is characterized by two parameters,
the inductance $L$ and the capacitance $C$. Defining
$Z_b=\sqrt{L/C}$ and following the conventions of
Ref.~\onlinecite{devoret} the $\phi$-susceptibility is
\begin{equation}\label{suskis2}
\chi(\omega)=\frac{i}{\hbar}\int_0^{\infty}e^{i\omega
t}\langle\left[\phi(t),\phi(0)\right]\rangle dt,
\end{equation}
where
\begin{equation}\label{phi}
\phi=\left(\frac{\hbar Z_b}{2}\right)^{\frac{1}{2}}(a+a^{\dagger}).
\end{equation}
 The conjugate  of $\phi$ is
the charge operator
\begin{equation}\label{charge}
q=i\left(\frac{\hbar}{2Z_b}\right)^{\frac{1}{2}}(a^{\dagger}-a).
\end{equation}
According to the linear response theory, the susceptibility is
proportional to the impedance of the system,
$Z(\omega)=i\omega\chi(\omega)$.

To generalize $\chi(\omega)$ to nonsteady-state cases, we also
define a time-dependent susceptibility as
\begin{equation}\label{suskis3}
\chi(\omega,t)=\frac{i}{\hbar}\int_0^{\infty}e^{i\omega
(t'-t)}\langle\left[\phi(t'),\phi(t)\right]\rangle\,d(t'-t).
\end{equation}
When relaxation caused by the environment is weak,
$\chi(\omega,t)$ changes slowly compared with the internal
dynamics of the system and may be considered as a quasistatic
quantity. This is the case with a high $Q$ factor system. In a
nonsteady state the emission and absorption between the system
energy levels does not obey the detailed balance condition and
$\chi(\omega,t)$ may differ qualitatively from the steady-state
susceptibility $\chi(\omega)$ studied in Ref.~\onlinecite{rau}.
However, the oscillator-qubit system approaches the thermal
equilibrium irrespective of the initial state and, therefore,
\begin{equation}\label{relax}
\lim_{t\to\infty}\chi(\omega,t)=\chi(\omega).
\end{equation}
In practice, $\chi(\omega,t)\approx\chi(\omega)$ for $t \gtrsim
Q/\omega_0$.

\begin{figure}[h]
\centering
\begin{picture}(200,80)
\put(-15,20){\includegraphics[width=0.9\columnwidth]{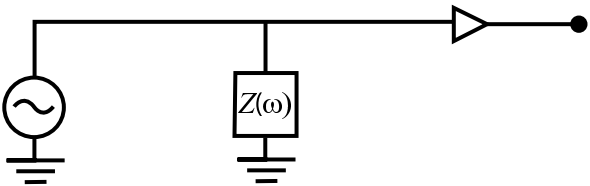}}
\end{picture}
\caption{ Realization to carry out transmission impedance
measurement of a solid-state quantum circuit. The impedance
$Z(\omega)$ represents the coupled oscillator-qubit system. The
impedance $Z(\omega)$ is dependent on the quantum state of the
system which can be probed by a microwave pulse.} \label{imp}
\end{figure}

The susceptibility function contains knowledge about the
dissipative and reactive response of the system under a weak
external perturbation. The susceptibility at a certain moment
$\chi(\omega,t_0)$ can be measured by exposing the system to a
perturbation acting rapidly compared to the natural relaxation
time of the system. In the solid-state context this could be
carried out by connecting the system of interest between input and
output transmission lines (Fig.~\ref{imp})
 and measuring the
transmission of a microwave pulse. Different quantum states of the
coupled oscillator and qubit system result in different impedances.
By measuring the voltage of the transmitted signal one can acquire
knowledge about the qubit state. The most important requirement for
the single-shot measurement of this kind is that the mixing due to
the microwave drive and the relaxation caused by the environment
should take much longer than the voltage measurement necessary to
identify the state of the system.
Assuming the drive-induced mixing to be significantly quicker than
the environment-induced relaxation, one can calculate a simple
estimate for the mixing time. The driving Hamiltonian is assumed to
take the form
\begin{equation}\label{drive}
H_d=V_0\,q\,\mathrm{cos}(\omega_d \,t),
\end{equation}
where $q$ is the charge operator (\ref{charge}) and $V_0$ describes
the strength of the voltage induced by the microwave field. We
assume that the frequency of the measurement signal $\omega_d$
coincides closely with an energy difference of two arbitrary energy
eigenstates $|i\rangle$ and $|f\rangle$ of the system. An
application of the first order time-dependent perturbation theory
yields an approximate transition time
\begin{equation}\label{mixing}
T_d\approx\frac{2\hbar}{V_0|\langle
f|q|i\rangle|}=\frac{\hbar}{eV_0}\sqrt{\frac{Z_b}{R_Q}}\frac{4\sqrt{\pi}}{|\langle
f | \hat{p} | i\rangle},
\end{equation}
where $R_Q=h/e^2 \approx 25.6{\mathrm k}\Omega$ and
$\hat{p}=i(\daka{a}-a)$. After time $T_d$ the microwave pulse has
disturbed the system and lost the information about the initial
state in $\chi(\omega_d,t_0)$.

The measurement time $T_m$ is limited from below by the voltage
noise of the output signal.  The measurement time can be estimated
by the signal-to-noise formula
\begin{equation}\label{meas}
S/N=\frac{\Delta V}{\sqrt{S_VT_m^{-1}}},
\end{equation}
where $S_V$ is the spectral density of the voltage fluctuations
and $\Delta V$ is the voltage difference in the signal caused by
two differing impedances $Z_1(\omega)$ and $Z_2(\omega)$ of the
oscillator. The different impedances $Z_1(\omega)$ and
$Z_2(\omega)$ correspond to two different quantum states of the
oscillator-qubit system. Supposing that the dominant voltage
fluctuations originate from the amplifier, we can write $S_V=2
k_{\mathrm{B}}\,T_N\,Z_0$, where $T_N$ is the noise temperature of
the amplifier and $Z_0$ is the characteristic impedance of the
transmission line. Setting $S/N=1$ and solving $T_m$ from
Eq.~(\ref{meas}) we get
\begin{equation}\label{measur}
T_m=\frac{2k_{\mathrm{B}}\,T_N\,Z_0}{(\Delta V)^2}=4\pi\frac{\hbar
k_B T_N}{(e\Delta V)^2}\frac{Z_0}{R_Q}.
\end{equation}
The transmitted voltage difference between two different states is
\begin{equation}\label{ero}
\Delta V=|V_{1}-V_{2}|=V_0 z,
\end{equation}
where we defined
\begin{equation}\label{transker}
z \equiv \left|\frac{2(Z_{1}/Z_0-Z_{2}/Z_0)}
{(2Z_{1}/Z_0+1)(2Z_{2}/Z_0+1)}\right|.
\end{equation}
The voltage difference is significant when $Z_{1}$ and $Z_{1}$ are
of order $Z_0$ or one is much larger and the other much smaller
than $Z_0$.

For carrying out a transmission measurement capable of resolving
two quantum states the condition $T_d\gg T_m$ must be fulfilled.
Using Eqs. (\ref{mixing}) and (\ref{measur}) their ratio is
\begin{equation}\label{suhde}
 \frac{T_{\mathrm{d}}}{T_{\mathrm{m}}}=
 \frac{\hbar\,V_0\,z^2}{k_{\mathrm{B}}T_N Z_0|\langle f|q|i\rangle|} = \frac{eV_0}{k_B T_N} \frac{\sqrt{Z_b R_Q}}{\sqrt{\pi} Z_0} \frac{z^2}{|\langle f|\hat{p}|i\rangle|},
\end{equation}
In the analogous case of current driving \cite{drivingnote}, the
ratio between the relaxation and measurement times is
\begin{equation}\label{suhde2}
 \frac{T_{\mathrm{d}}}{T_{\mathrm{m}}}=
 \frac{\hbar\,I_0\,z^2 Z_0}{k_{\mathrm{B}}T_N |\langle f|\phi|i\rangle|} = \frac{\frac{\hbar}{e} I_0}{k_B T_N} \frac{Z_0}{\sqrt{Z_b R_Q}} \frac{2\sqrt{\pi} z^2}{|\langle f|\hat{x}|i\rangle|},
\end{equation}
where $\hat{x}=\daka{a}+a$ and $I_0$ is the amplitude of the
driving field. The matrix element $|\langle f|\hat{p}|i\rangle|$
in Eq.~(\ref{suhde}) and $|\langle f|\hat{x}|i\rangle|$ in
Eq.~(\ref{suhde2}) are of the order of unity in transitions
between the lowest energy states. The impedance $Z_0$ can be
matched with $Z_1$ or $Z_2$ via a matching circuit, in order to
produce an optimal signal. If the requirement $T_d \gg T_m$ is not
met, one needs to perform a sequence of measurements to resolve
the qubit state.

\section{Results} \label{results}

Let us now turn to the impedance
$Z(\omega,t)=i\,\omega\chi(\omega,t)$ of the strongly coupled
qubit-oscillator system. We characterize the different qubit
states by using impedance information and apply this knowledge to
the qubit measurements.

We illustrate the dependence of the oscillator impedance on the
qubit state by considering an example. In the following numerical
results, we choose the coupling strength $g=0.03\omega_0$ between
the oscillator and the qubit and the quality factor $Q=10^4$ of
the oscillator. These parameter values are accessible in the
solid-state realizations \cite{blais}. The ratio between the
mixing times $\gamma$ and $\kappa$ in the qubit and the
oscillator, respectively, is chosen to be $\gamma/\kappa=0.08$.
This in agreement with Ref.~\onlinecite{blais}

\begin{figure}
\begin{center}
\begin{picture}(200,180)
\put(-0,10){\includegraphics[width=0.9\columnwidth]{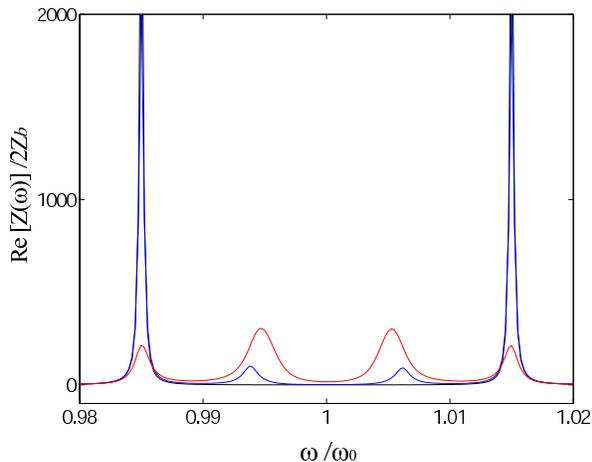}}
\end{picture}
\caption{Real part of impedance at $T=\hbar\omega_0/k_{\mathrm{b}}$
(red), $T=\hbar\omega_0/3k_{\mathrm{b}}$ (blue) and
$T=\hbar\omega_0/10k_{\mathrm{b}}$ (black).} \label{kuva1}
\end{center}
\end{figure}
\begin{figure}
\begin{center}
\begin{picture}(200,180)
\put(-0,10){\includegraphics[width=0.9\columnwidth]{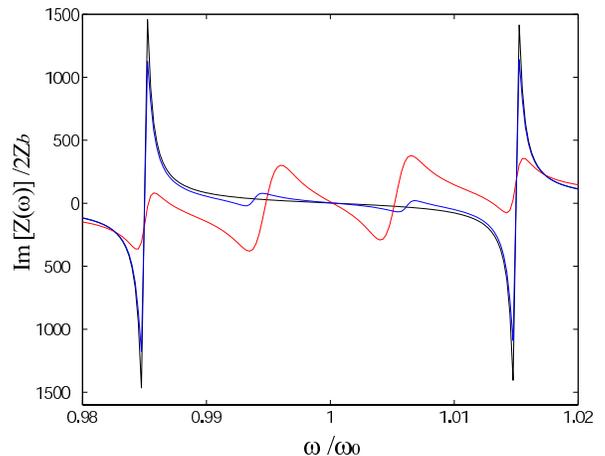}}
\end{picture}
\caption{Imaginary part of the equilibrium impedance at
$T=\hbar\omega_0/k_{\mathrm{b}}$ (red),
$T=\hbar\omega_0/3k_{\mathrm{b}}$ (blue) and
$T=\hbar\omega_0/10k_{\mathrm{b}}$ (black).} \label{kuva2}
\end{center}
\end{figure}

The real and imaginary parts of the impedance $Z$ of the system
are shown in Figs.~\ref{kuva1} and \ref{kuva2} for three different
temperatures $T$ of the bath. The coupling between the oscillator
and the qubit results in multiple peaks in strong contrast to a
single oscillator. Moreover, the impedance of the coupled system
is now strongly dependent on temperature. In the low temperature
limit there are only two peaks in ${\rm Re}[Z(\omega)]$,
corresponding to the vacuum Rabi splitting. They have been
experimentally observed recently for example in
Ref.~\onlinecite{wallraff}. The susceptibility is studied as a
function of temperature in detail in Ref.~\onlinecite{rau}.

\begin{figure}[h]
\begin{picture}(200,400)
\put(10,200){\includegraphics[width=0.9\columnwidth]{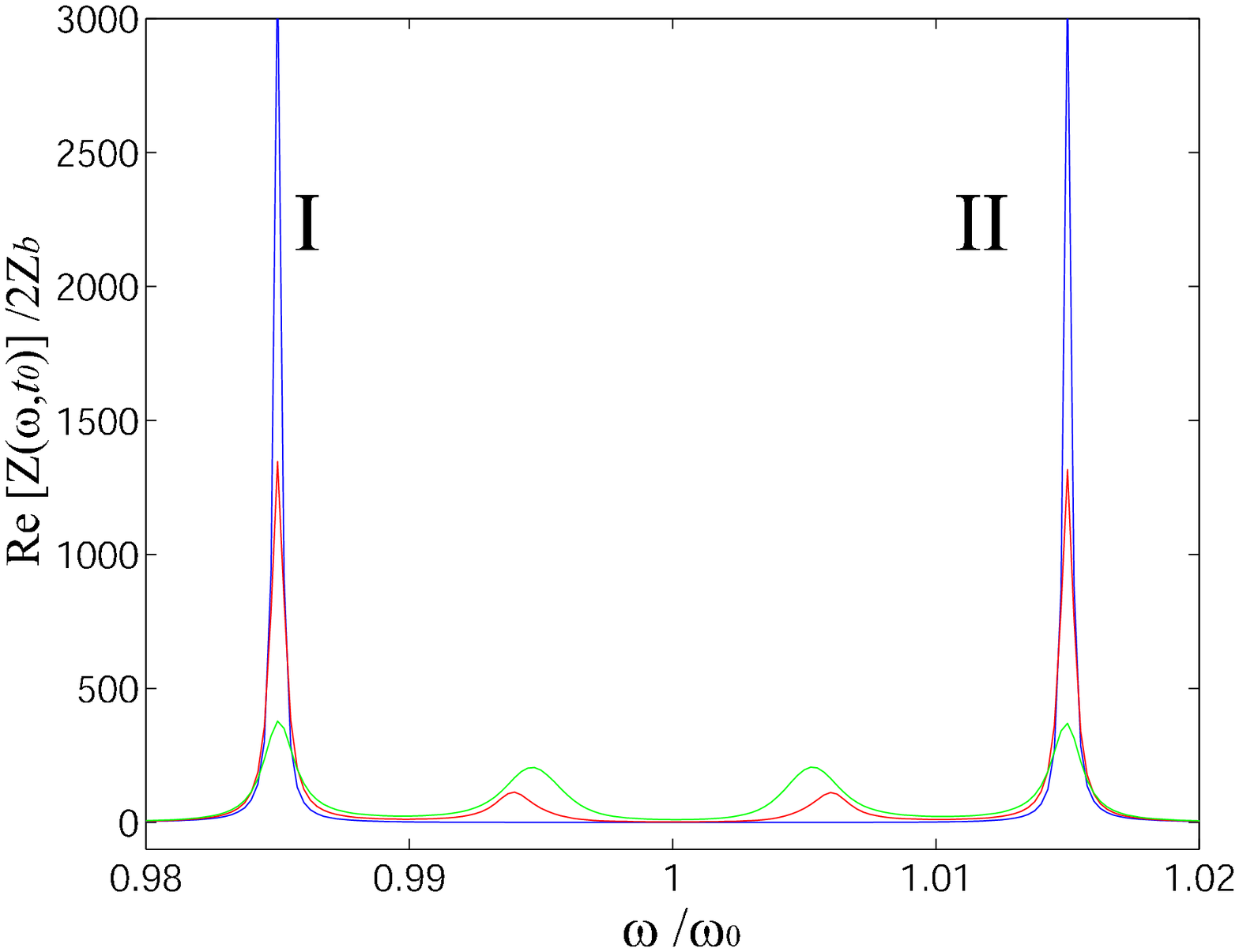}}
\put(10,10){\includegraphics[width=.9\columnwidth]{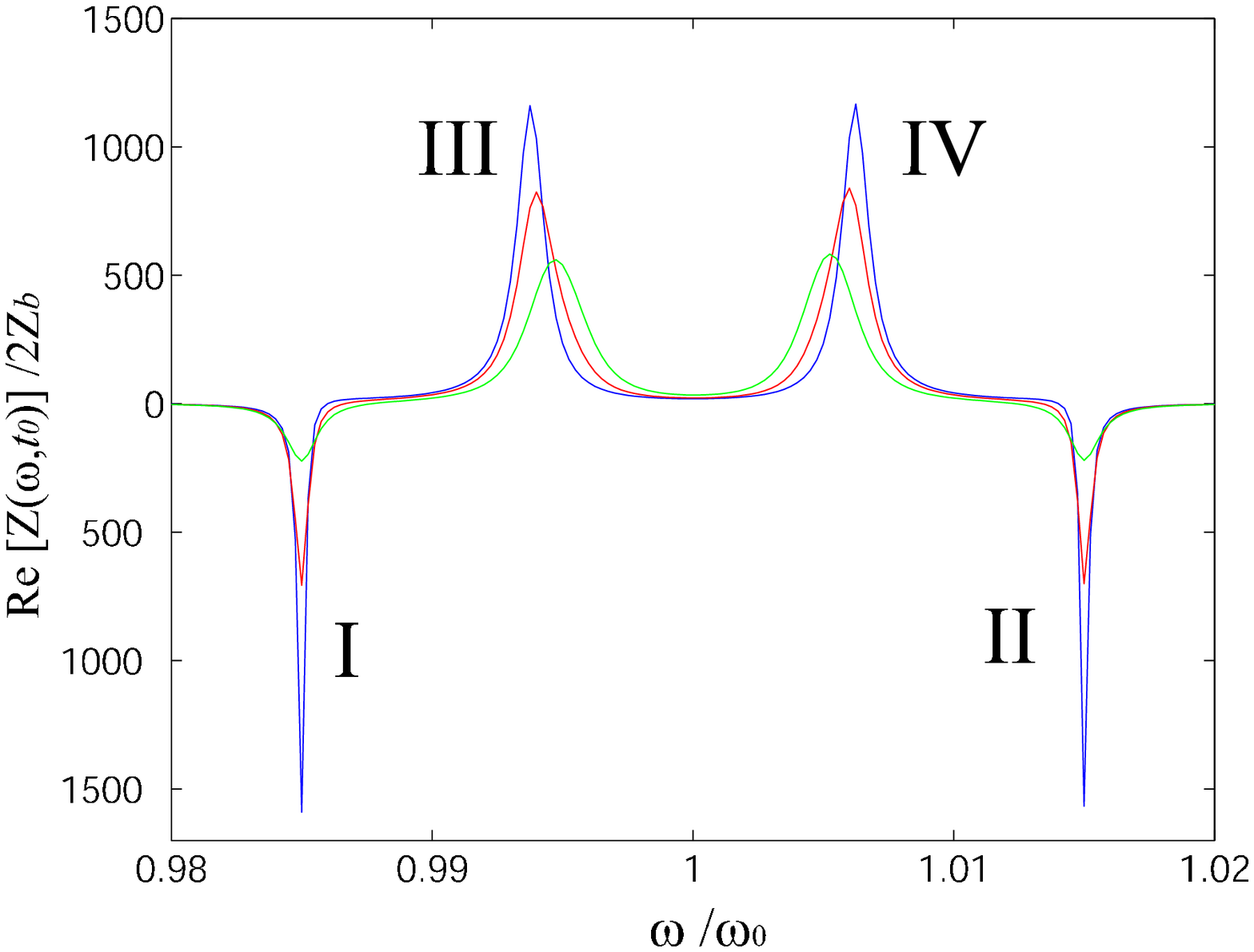}}
\end{picture}
\caption{(a) Real part of $Z(\omega,t_0)$ at temperatures
$T=\hbar\omega_0/10k_{\mathrm{b}}$ (blue),
$T=\hbar\omega_0/2k_{\mathrm{b}}$ (red) and
$T=\hbar\omega_0/k_{\mathrm{b}}$ (green). The initial state at
$t=t_0$ is prepared so that the oscillator is in the thermal state
and the qubit state is up (the lower energy qubit state). (b) Same
as (a) but the initial state is prepared so that the qubit state
is down.}\label{ensim}
\end{figure}

When the oscillator-qubit system is not in thermal equilibrium, the
susceptibility changes radically (see Fig.~\ref{ensim}). The
detailed-balance condition, which relates emission and absorption
processes, is violated. For example, one can have net emission of
energy in some frequencies in addition to the absorption. In those
frequencies the susceptibility takes negative values which lead to
negative impedance peaks as in Fig.~\ref{ensim}(b). The negative
peaks are signs of spontaneous emission of energy and relaxation
towards the equilibrium state.

\begin{figure}[h]
\centering
\begin{picture}(200,130)
\put(15,5){\includegraphics[width=0.6\columnwidth]{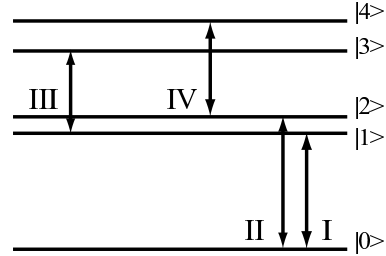}}
\end{picture}
\caption{Transitions corresponding to the four peaks of the low
temperature impedance curves (blue) in Fig.~\ref{ensim}(a) and
(b). The transitions I and II correspond to the vacuum Rabi
splitting in (a). The reason that the peaks I and II are negative
in Fig.~\ref{ensim}(b) is that the initial state is far from
equilibrium and spontaneously emits energy.} \label{etaso}
\end{figure}

Both sets of impedance curves in Fig.~\ref{ensim} (a) and (b)
contain characteristic peaks which can be utilized to resolve the
state of the qubit. At low temperatures the peaks III and IV are
absent in Fig.~\ref{ensim} (a). In addition, peaks I and II are
negative in Fig.~\ref{ensim} (b). The transitions I-IV are shown in
Fig. \ref{etaso}. They correspond to the transitions between the
lowest energy states of the coupled system. When the coupling $g$
between the oscillator and the qubit is weak, the lowest energy
states are given in terms of the uncoupled eigenstates as
\begin{equation*}
\begin{split}
|0\rangle&=|n=0\rangle\: |\uparrow\rangle\nonumber\\
|1\rangle&=\frac{1}{\sqrt{2}}\big(|n=0\rangle\: |\downarrow\rangle+|n=1\rangle\: |\uparrow\rangle \big)\\
|2\rangle&=\frac{1}{\sqrt{2}}\left(|n=0\rangle\: |\downarrow\rangle-|n=1\rangle\: |\uparrow\rangle\right)\\
|3\rangle&=\frac{1}{\sqrt{2}}\left(|n=1\rangle\: |\downarrow\rangle+|n=2\rangle\: |\uparrow\rangle\right)\\
|4\rangle&=\frac{1}{\sqrt{2}}\left(|n=1\rangle\:
|\downarrow\rangle-|n=2\rangle\: |\uparrow\rangle\right).
\end{split}
\end{equation*}
The reason for the negative peaks in Fig.~\ref{ensim} (b) is that
the ground state is unpopulated in the initial state and the system
begins to spontaneously increase the ground state population. All
peaks that remain present even in the zero temperature limit are
higher and narrower at lower temperatures. At higher temperatures
new peaks arise. Comparing curves Fig.~\ref{ensim} (a) and
Fig.~\ref{ensim} (b) one can also notice that the growth of
temperature decreases the differences between the
 impedance curves corresponding to the
different states of the qubit. This phenomenon restricts the
accuracy of the qubit measurement based on the impedance
difference and puts an upper limit to the measurement temperature
to roughly $T \sim \hbar \omega_0/k_b$.

\begin{figure}
\begin{center}
\begin{picture}(200,198)
\put(10,10){\includegraphics[width=0.8\columnwidth]{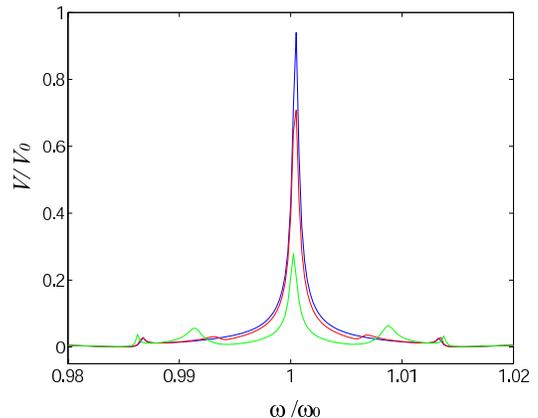}}
\end{picture}
\caption{Difference between the transmitted current amplitudes
corresponding to the qubit states up and down when the oscillator
is in thermal equilibrium at $T=\hbar\omega_0/10k_{\mathrm{b}}$
(blue), $T=\hbar\omega_0/4k_{\mathrm{b}}$ (red) and
$T=\hbar\omega_0/2k_{\mathrm{b}}$ (green). The ratio of the
impedances $Z_b$ and $Z_0$ is set to $Z_0/Z_b=2$. }
\label{transmiss}
\end{center}
\end{figure}

\begin{figure}
\begin{center}
\begin{picture}(200,198)
\put(10,10){\includegraphics[width=0.8\columnwidth]{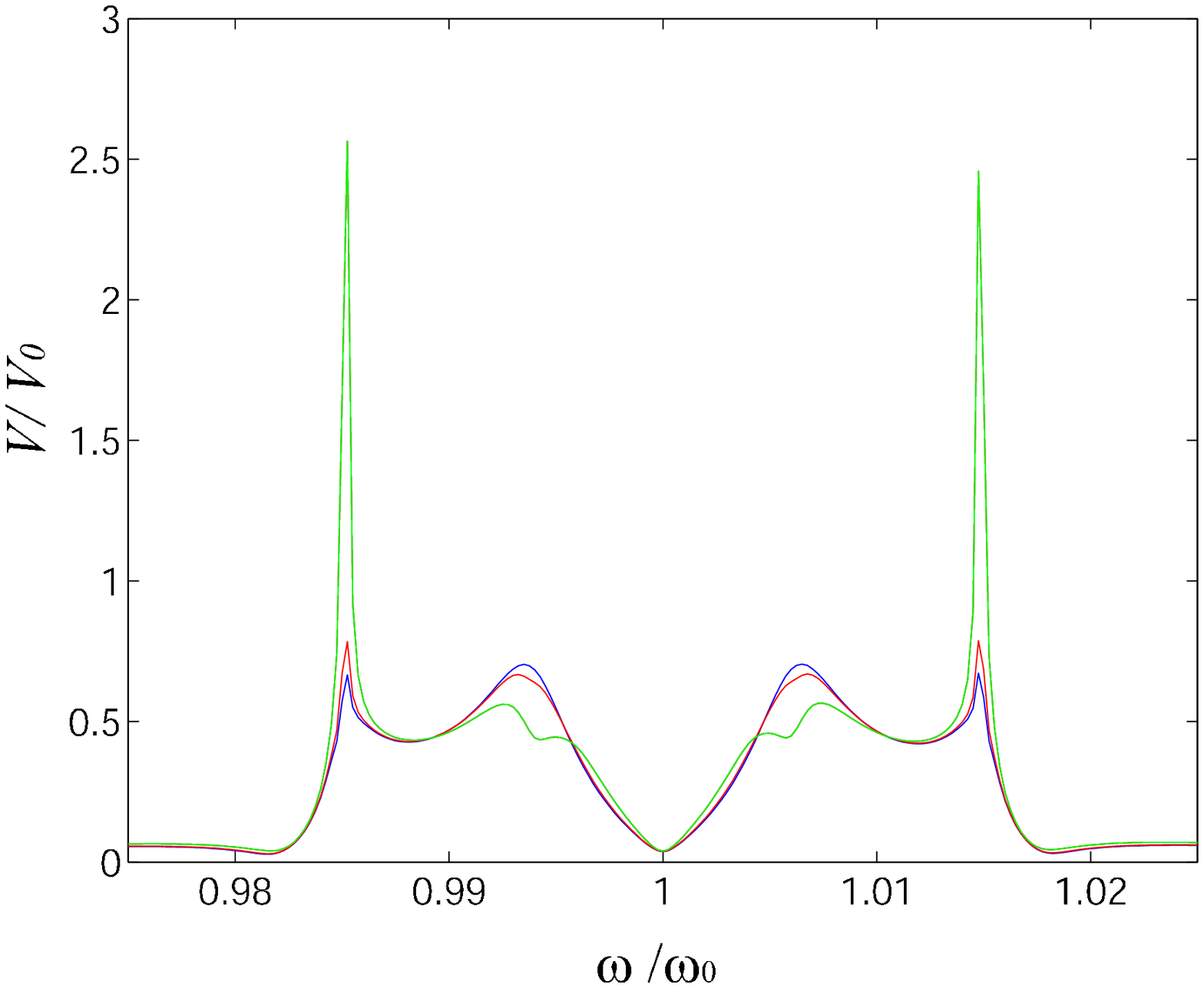}}
\end{picture}
\caption{Same as Fig.~\ref{transmiss} but with voltage driving
(impedances replaced by admittances, and calculated from the
$q$-susceptibility) and $Z_0/Z_b=1/500$.} \label{transmiss1}
\end{center}
\end{figure}

In Figs.~\ref{transmiss} and \ref{transmiss1} we have plotted the
difference in the transmitted voltage amplitudes (\ref{ero}),
corresponding to the current ($\phi$-susceptibility) and voltage
($q$-susceptibility) driving, respectively. In
Fig.~\ref{transmiss} the maximum difference do not coincide with
the peaks I-IV. This is because the impedance of the system at the
peaks is much larger than the characteristic impedance of the
transmission line $Z_0$ and the system effectively decouples from
it. However, using an appropriate matching circuit the effective
impedance of the system can be varied. Better matching is obtained
in Fig.~\ref{transmiss1} where we consider voltage driven system
and the impedance ratio is chosen to $Z_0/Z_b=1/500$. The
transmitted voltage amplitude can exceed unity because of the
emission associated with the transition I and II. At temperatures
$T=\hbar\omega_0/10k_{\mathrm{b}}$ and
$T=\hbar\omega_0/4k_{\mathrm{b}}$ the absolute values of the
impedances at the peaks I and II are so high that the system again
decouples from the transmission line. Thats why the transmitted
voltage corresponding to $T=\hbar\omega_0/2k_{\mathrm{b}}$ gives
the maximum for this particular impedance matching. The curves are
not quite symmetrical with respect to $\omega_0$ because the
imaginary part of the impedance is antisymmetric with respect to
$\omega_0$.


\begin{figure}
\begin{center}
\begin{picture}(200,198)
\put(10,10){\includegraphics[width=0.8\columnwidth]{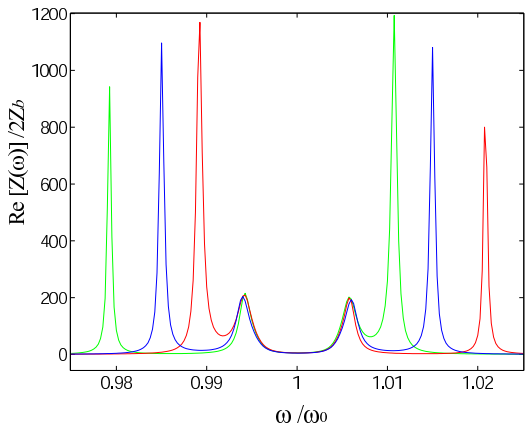}}
\end{picture}
\caption{Equilibrium impedance of a resonant and a slightly
off-resonant oscillator-qubit system at
$T=\hbar\omega_0/2k_{\mathrm{b}}$. The blue curve represents the
exactly resonant case, the red curve corresponds to the case
$\omega_{\mathrm{qb}}=1.01\,\omega_0$ and the green curve to the
case $\omega_{\mathrm{qb}}=0.99\,\omega_0$. } \label{nonres}
\end{center}
\end{figure}

 The impedance
of a slightly off-resonant oscillator-qubit system is plotted in
Fig.~\ref{nonres}. The vacuum Rabi splitting, corresponding to the
transitions I and II, is sensitive even to a slight detuning and
thus the positions of these peaks follow the frequency
$\omega_{qb}$ of the qubit. On the other hand, the positions of
the peaks corresponding to the transitions III and IV follow
rather the frequency $\omega_0$ of the oscillator. With small
detuning, the height of the impedance peaks is only slightly
altered.

\begin{figure}
\begin{picture}(200,600)
\put(0,430){\includegraphics[width=0.8\columnwidth]{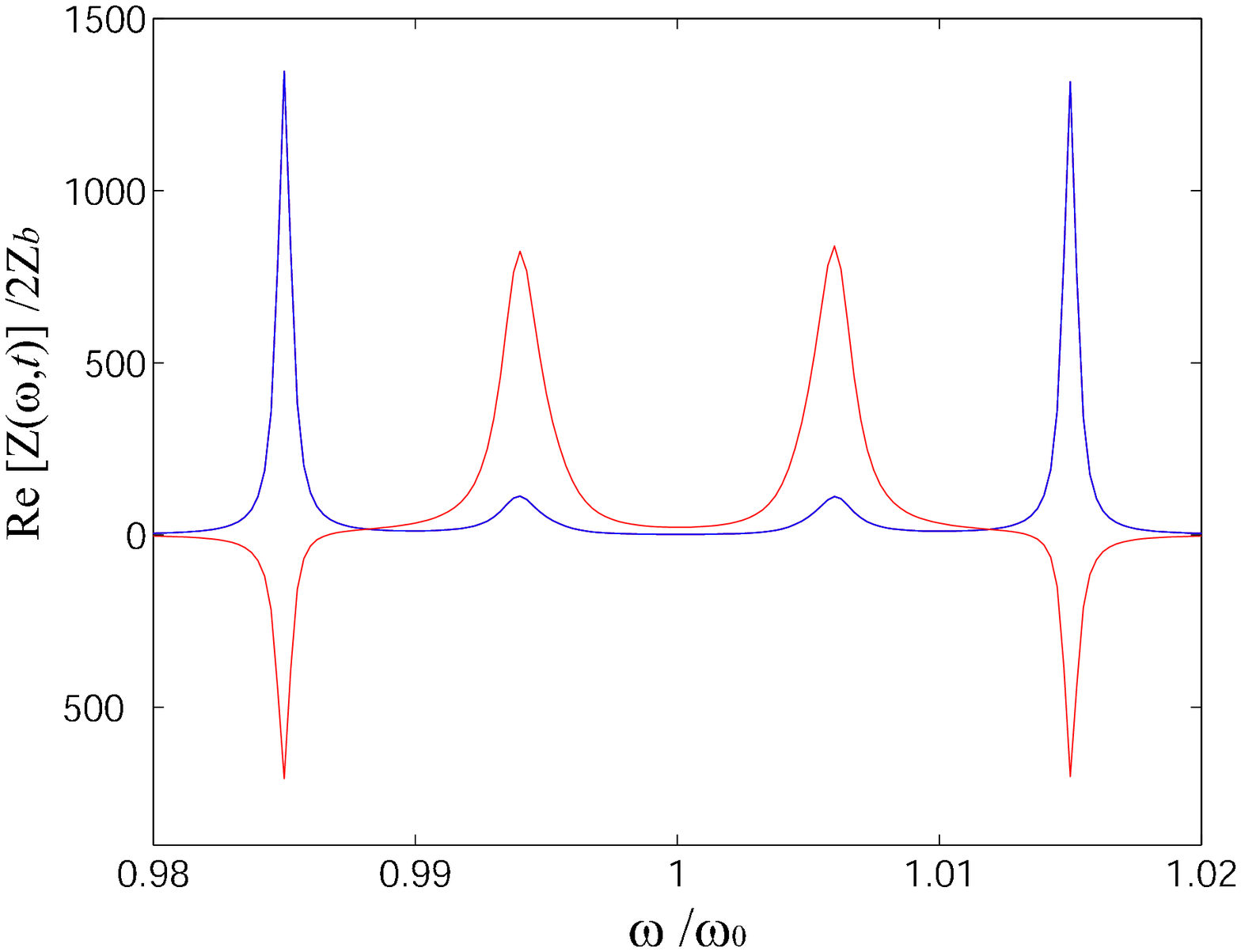}}
\put(0,230){\includegraphics[width=.8\columnwidth]{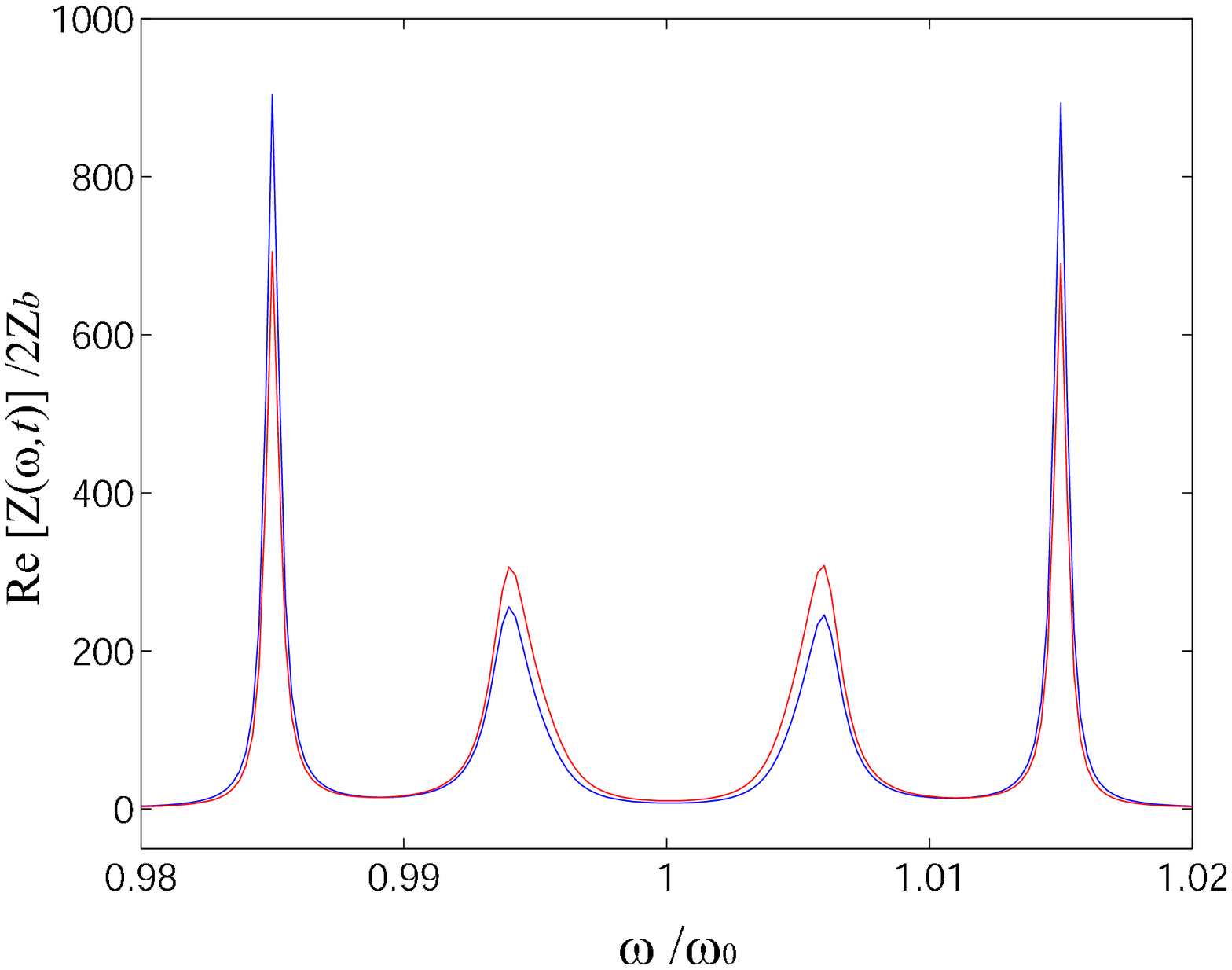}}
\put(0,30){\includegraphics[width=.8\columnwidth]{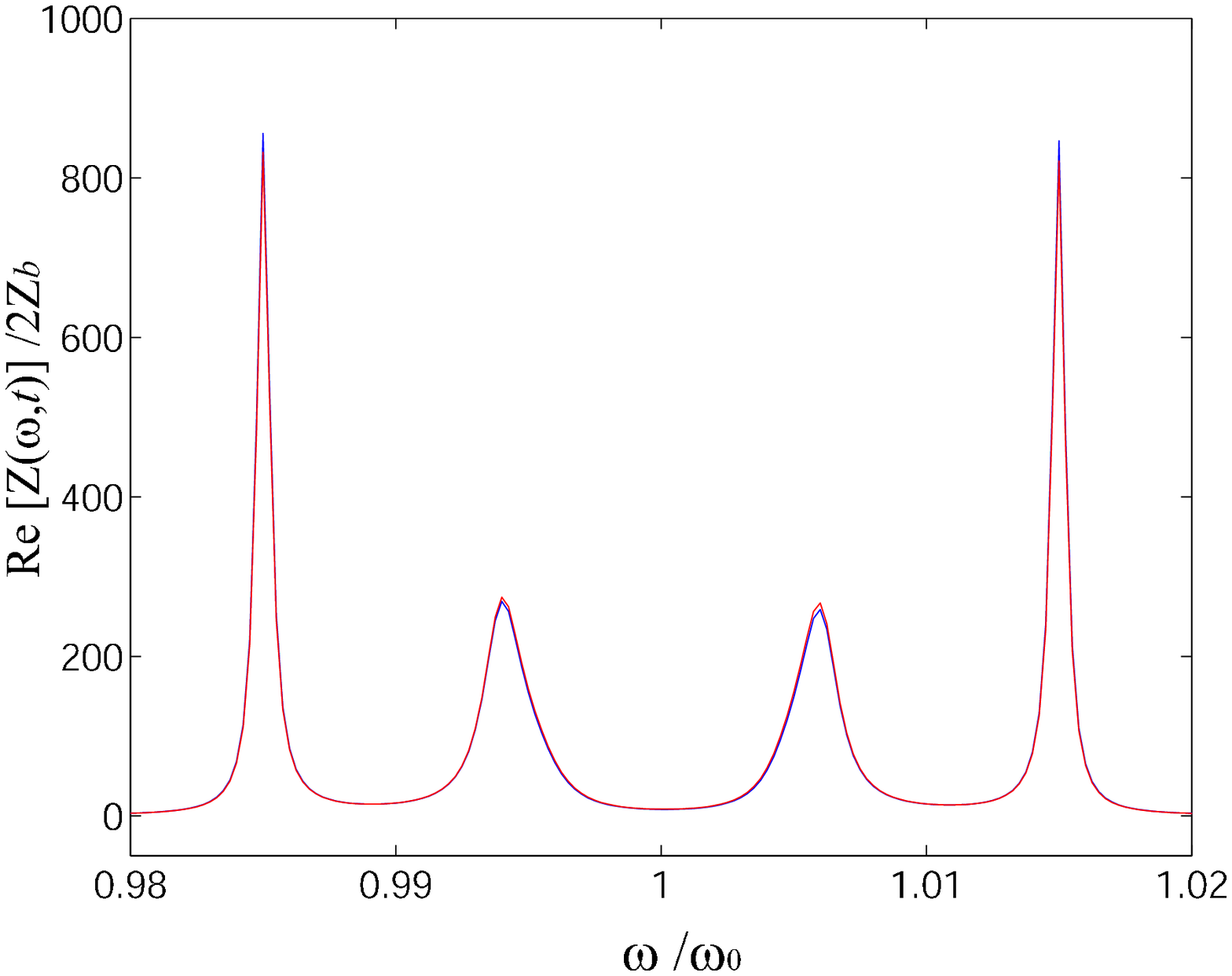}}
\end{picture}
\caption{Relaxation of the impedance at
$T=\hbar\omega_0/1.8k_{\mathrm{b}}$ as a function of time. The
initial state at $t=0$ is prepared so that the oscillator is in
the thermal state and the qubit state is up (red line) or down
(blue line). The figures correspond to the moments $t=0$,
$t=5000\,\omega_0^{-1}$ and $t=10000\,\omega_0^{-1}$.
}\label{toinen}
\end{figure}

As the system relaxes towards the equilibrium state, the impedance
settles to the equilibrium pattern. In Fig.~\ref{toinen} we plot
the temporal evolution of two nonequilibrium impedances. They
correspond to the initial states where the oscillator is in
thermal equlibrium and the qubit is prepared either up or down.
The susceptibility changes slowly compared to $\omega_0^{-1}$ and
can be considered as quasistatic. After the time $t\gtrsim
Q/\omega_0$ the susceptibilities in both cases are nearly equal,
reflecting the uncertainty about the state of the qubit
(Fig.~\ref{toinen}). When $\gamma \ll \kappa$, the oscillator
dissipation yields the dominant time scale for the relaxation of
the qubit near resonance.

\section{Conclusions}

In this paper we discuss the susceptibility and the impedance of a
resonant oscillator-qubit system at finite temperatures. We have
studied the response of the system in various nonequilibrium
states and explored possibilities to apply this phenomenon to
determine the state of the qubit. In certain conditions one can
carry out a transmission measurement capable of resolving the
qubit state.

Let us estimate the time scales $T_d$ and $T_m$ for an example
case with voltage driving. Assume the parameters for the
oscillator to be $L=200$nH and $C=50$fF. This yields the resonant
angular frequency $\omega_0=10$GHz and characteristic impedance
$Z_b=2$k$\Omega$. Suppose that we are performing the transmission
measurement at $T=\hbar \omega_0/(4 k_b)=20$mK or $T=\hbar
\omega_0/(2 k_b)=40$mK, corresponding to the red and green curves
in Fig.~\ref{transmiss1}. We further assume that the amplifier is
impedance matched such that the amplifier looks to the sample as
an impedance $Z_0=Z_b/500$. This can be accomplished via a
transformer circuit. Now taking $eV_0=0.03 \hbar \omega_0=0.2\mu$V
and $T_N=5\hbar \omega_0/k_b=400$mK, we get the measuring time
$T_m=17/\omega_0=1.7$ns, and the relaxation time
$T_d=93/\omega_0=9.3$ns. Thus, it should be possible to extract
the information of the state of the qubit in a single measurement.
These numbers were calculated by assuming $z=0.8$ and $|\langle
f|\hat{p}|i\rangle=1/\sqrt{2}$. The ratio $T_d/T_m$ can be further
improved for voltage driving if $Z_b$ can be made larger, $Z_0$
smaller, or if the noise temperature $T_N$ can be decreased. Note
that the estimate (\ref{mixing}) is a pessimistic one: the
off-resonant mixing time is longer than when the driving frequency
corresponds exactly to the position of the impedance peaks.


\section*{Acknowledgement}
We thank Mika Sillanp\"a\"a, Pertti Hakonen and G\"oran Johansson
for valuable discussions. TTH acknowledges the financial support
by the Academy of Finland and the NCCR Nanoscience.


\begin{thebibliography}{10}
\bibitem{drivingnote} The case of voltage driving is realized if
the inductor $L$ and capacitor $C$ are in series, and of current
driving for their parallel configuration. In the previous case,
one is interested in the $q$-susceptibility, and in the latter
case, in the $\phi$-susceptibility. The results for the two are
qualitatively similar, in the former case the admittance taking
the role of the impedance.
\end{thebibliography}
\end{document}